\documentstyle[aps,epsf]{revtex}
\begin{document}
\draft
\preprint{ }
\title{The Surface of a Bose-Einstein Condensed
Atomic Cloud}
\author{U. Al Khawaja$^{1}$,  C. J. Pethick$^{2}$ and H. Smith$^{1}$}
\address{$^1$\O rsted Laboratory, H. C. \O rsted Institute,
Universitetsparken 5, DK-2100 Copenhagen \O, Denmark\\
$^2$Nordita, Blegdamsvej 17, DK-2100 Copenhagen \O, Denmark.\\
$^3$Department of Physics, University of Illinois at Urbana-Champaign,
1110 W.\ Green Street, Urbana, Illinois 61801, USA.}

\date{\today}

\maketitle

\begin{abstract}
We investigate the structure
and  collective modes of a planar surface of a trapped Bose-Einstein
condensed gas  at zero temperature.
In the long-wavelength limit we find a mode similar to the
gravity wave on the surface of a fluid with the frequency $\omega$ and 
the wavenumber
$q$ related by $\omega^2 =Fq/m$. Here $F$ is the force due to
the confining potential
at the surface and $m$ is the particle mass.
At shorter wavelengths we use a variational approach and find
corrections to $\omega^2$ of order $q^4 \ln{q}$.
We demonstrate the usefulness of the concept of an effective surface tension
for describing both  static and dynamic properties of condensed
atomic clouds.

   \end{abstract}


\section{Introduction}

Properties of trapped clouds of Bose-Einstein condensed atoms have been
investigated  intensively both experimentally
and theoretically over
the past few years, since the experimental realization
of Bose-Einstein condensation in dilute atomic gases\cite{123}.
At zero temperature, the behavior of the order parameter, $\psi$,
is determined by the
time-dependent Gross-Pitaevskii equation, a 
Schr\"{o}\-dinger equation with
the nonlinear term proportional to $|\psi|^2\psi$ added:
\begin{equation}
- \frac{\hbar^2}{2m}\nabla^2 \psi({\bf r},t)
 +V({\bf r})\psi({\bf r},t)
 +U_0|\psi({\bf r},t)|^2\psi({\bf r},t)=i\hbar
\frac{\partial \psi({\bf r},t)}{\partial t}.
\label{gp}
\end{equation}
Here $U_0$ is the effective two-particle interaction,
which may be expressed in terms of the scattering length $a$
according to $U_0=4\pi a\hbar^2/m$, where $m$ is the mass of an atom.
The confining potential is denoted by $V({\bf r})$.
The static structure of the cloud is determined by the time-independent
Gross-Pitaevskii
equation, which reads
\begin{equation}
- \frac{\hbar^2}{2m}\nabla^2 \psi({\bf r})
 +V({\bf r})\psi({\bf r})
 +U_0|\psi({\bf r})|^2\psi({\bf r})=\mu \psi({\bf r}),
\label{staticgp}
\end{equation}
where $\mu$ is the chemical potential.  For clouds containing a sufficiently
large number of atoms with repulsive interactions,
many properties may be calculated to a good approximation using
the Thomas-Fermi approximation\cite{ruckenstein leggett bp}, in which one
neglects the kinetic energy term in the Gross-Pitaevskii equation.
One then finds a density profile
\begin{equation}
|\psi({\bf r})|^2 =(\mu-V({\bf r}))/U_0,\; \mu\ge V({\bf r}),\;\;
|\psi({\bf r})|^2 =0,\;                    \mu<V({\bf r}).
\end{equation}
The Thomas-Fermi approach may also be used for calculating collective
modes, and this has been done by Stringari\cite{stringari}.

A number of phenomena are
 associated with the surface region of the cloud.  One
is that the contribution of the kinetic energy term to the total energy of the
cloud comes mainly from the surface region.  Also Stringari\cite{stringari}
 has identified
surface modes of oscillation in the Thomas-Fermi approximation.  To understand
such phenomena in detail, it is useful to consider the properties of a planar
surface, and to approximate the trapping potential by
 a linear function of the coordinates.  The potential is then given by
\begin{equation}
V({\bf r})= Fx,
\end{equation}
where the coordinate $x$ measures distances in the direction  of
$\nabla V$.  The presence of
the $\nabla^2$ term in the Gross-Pitaevskii equation
leads to a rounding-off of the density profile in the surface region over a
distance of order
\begin{equation}
\delta= \left(\frac{\hbar^2}{2mF}\right)^{1/3},
\label{delta}
\end{equation}
and within this
approach the  static structure of the surface region has been studied,
and contributions to
the total energy  calculated \cite{pitaevskii lundh fetter}.

The purpose of this paper is to consider   properties of plane
surfaces of Bose-Einstein condensed clouds in a linear potential.  This is of
interest both for giving analytical expressions for a number of properties of
large clouds, as well as for giving physical insight.
 First, we consider the
static properties of the surface region, and show how the kinetic energy
term in the Gross-Pitaevskii equation gives rise to an effective surface
tension.  We also consider the density profile in the vicinity of the surface,
and show that the Gross-Pitaevskii equation and the Thomas-Fermi
approximation to it lead to identical results for the column density of atoms
down to a point well within the cloud.  In the second part of the paper we
consider surface collective modes.  We begin by considering the Thomas-Fermi
approach, which yields a mode with the dispersion
relation $\omega^2=(F/m)q$, where
$\omega$ is the angular frequency and $q$ the magnitude of the wave vector
for the mode. This is
exactly the same form as for gravity waves on the surface of a fluid, with the
role of gravity being played by the trapping potential.  At higher wave
numbers
there are contributions to the mode
frequency arising from the $\nabla^2\psi$
term in the Gross-Pitaevskii equation, and we evaluate the effect of these by
employing a trial wave function that allows for the rounding-off of the order
parameter profile in the surface region.  The extra contributions to
$\omega^2$ are proportional to $ q^4 \ln(1/ q\delta)$,
and they will
be shown to have a ready interpretation in terms of the effective surface
tension introduced in the discussion of static properties.

\section{Static properties}

In this section we introduce the concept of an effective surface tension, and
consider the number of particles that are associated with the rounding-off of
the density profile in the surface region.
We shall take the origin of the $x$-coordinate  to be at the point where the
potential is equal to the chemical potential, which is where the density
vanishes in the Thomas-Fermi approximation.

Let us now calculate the contribution to the energy
coming from the kinetic energy term in the Hamiltonian.
As was discussed in Ref. \cite{lps}, the {\it total} kinetic energy may
be written in terms of either $\int_{-\infty}^{\infty}{(\psi ')}^{2}dx$
or $\int_{-\infty}^{\infty}\psi \psi '' dx$,
where the prime denotes differentiation with respect to $x$:
the kinetic energy density is not a well defined quantity.
For calculating the total energy either expression may be used,
but it must be used consistently.
In the case of the surface problem under investigation, we wish to
be able to associate a contribution to the kinetic energy with the surface,
and there is no unambiguous way of doing this,
since one could use either of the energy expressions
integrated to some point well inside the surface
(rather than $-\infty$ as in the case of the total kinetic energy).
However, these differ by a constant.
For definiteness, we shall employ the symmetrized expression and define
the kinetic energy per unit area of the surface as
\begin{equation}
\epsilon_K = \frac{\hbar^2}{2m} \int_{-L}^{\infty} {(\psi ')}^2  dx,
\label{ke}
\end{equation}
 where the point $x=-L$ is chosen to lie well inside the surface.
 From Ref. \cite{lps} this may be seen to be given by
\begin{equation}
\epsilon_K \approx \frac{\hbar^2}{8m} \frac{F}{U_0}
\ln\left(\frac{L}{0.240\delta}\right),
\label{keapprox}
\end{equation}
where the coefficient 0.240 was found by numerical integration.
If we had chosen the other expression for the kinetic energy density,
the result would differ by $\hbar^{2}F/4m U_{0}$.
The coefficient of $\ln{L}$
is independent of the choice of kinetic energy density,
but terms independent of $L$ are not.
This difference, however, is unimportant in application to surface modes,
where the logarithmic term is the one of  most interest.
If this
expression were independent of $L$,
the kinetic energy would have precisely the same form as a
surface tension, since the energy  would be proportional to the
area of the surface. Because the kinetic energy density falls off
only slowly away from the surface, the total kinetic energy increases
logarithmically
with $L$, but this dependence is sufficiently mild that it is still
useful to think of the kinetic energy as being physically analogous to a
surface tension.

Next we consider the
density profile, and compare the Thomas-Fermi result with the
exact one.  For $|x|\gg \delta$ the Thomas-Fermi wave function is a good
approximation to the
exact one and thus $\psi''$ is negative.  Consequently, as one can see
from the Gross-Pitaevskii equation, the density is depressed below the
Thomas-Fermi value.  Outside the cloud, the density is increased, due to the
quantum-mechanical tail of the wave function.  We shall now demonstrate the
somewhat surprising result that the number of particles associated with the
depression of densities inside the cloud is exactly equal to the number
associated with regions of increased density.

To prove this result it is simplest to consider the momentum density, $\bf
g$, defined by
\begin{equation}
g_i =\frac{\hbar}{2i}(\psi^*\frac{\partial \psi}{\partial x_i}
-\psi\frac{\partial \psi^*}{\partial x_i}),
\label{mom}
\end{equation}
whose time dependence may be determined from the
 Gross-Pitaevskii equation (\ref{gp}). By differentiating (\ref{mom})
and using (\ref{gp}) we find that the  momentum density
satisfies the conservation condition
\begin{equation}
\frac{\partial g_i}{\partial t}+\frac{\partial T_{ij}}{\partial x_j}=
-n\frac{\partial V}{\partial x_i},
\label{momentum}
\end{equation}
 where $n (=|\psi|^2)$ is the density. Here the stress tensor
$T_{ij}$ is given by
\begin{equation}
T_{ij}=T_{ij}^0 + P\delta_{ij},
\label{stress}
\end{equation}
where
\begin{equation}
T_{ij}^0=\frac{\hbar^2}{2m}\left( \frac{\partial \psi}{\partial x_i}
\frac{\partial \psi^*}{\partial x_j}+\frac{\partial \psi^*}{\partial x_i}
\frac{\partial \psi}{\partial x_j}    -  \frac{1}{2} \frac{\partial^2
|\psi|^2}{ \partial
x_i \partial x_j}     \right)
\label{freestress}
\end{equation}
is the free-particle stress tensor, while $P$ is the pressure arising from the
interaction,
\begin{equation}
P=\frac{1}{2}n^2U_0.
\end{equation}
   For the one-dimensional problem we are
considering, the condition for hydrostatic equilibrium is simply
\begin{equation}
\frac{\partial T_{xx}}{\partial x} =-Fn,
\label{hydrostat}
\end{equation}
from which it follows that
\begin{equation}
T_{xx}(-L)=F\int_{-L}^{\infty}n dx,
\label{ab}
\end{equation}
since the stress tensor vanishes for
large positive $x$.  Deep within the cloud
the order parameter varies as $(-x)^{1/2}$, and therefore $T_{xx}^0(-L)\sim
1/L$, which vanishes for $L \rightarrow \infty$.  Consequently
\begin{equation}
\int_{-L}^{\infty}n dx = \frac{1}{2}n^2(-L) U_0 +{\cal O} (1/L).
\label{abc}
\end{equation}
The first term is the result one obtains in the Thomas-Fermi approximation,
where $n=-xF/U_0$ for $x<0$, and
thus one sees that the total column density outside a point well within the
surface is the same as that in the Thomas-Fermi approximation, apart from
corrections of order $1/L$.  Physically the result follows from the condition
that the stress tensor at any point must be balanced by the force on all the
material at larger values of $x$, a result familiar in the context of
equilibrium of fluids and gases in the presence of gravitational fields.

\section{Surface modes of oscillation at long wavelengths}

We turn now to time-dependent situations and consider oscillations. For this
purpose it is convenient
to use instead of (\ref{gp}) an equivalent set of
 equations for the density, and the local velocity of the condensate.
  The equations of motion are obtained by
writing $\psi $ in terms of its amplitude $f$
and phase $\phi$, $\psi=fe^{i\phi}$.
The number density is given by $n =f^2$,
while
 the velocity ${\bf v}$ is given by ${\bf v}=\hbar{\bf \nabla}\phi/m$.
By inserting  $\psi=fe^{i\phi}$ into (\ref{gp})
and separating the equation into real and imaginary parts
one  obtains the two equations
\begin{equation}
\frac{\partial (f^2)}{\partial t}
=-\frac{\hbar}{m}{\bf \nabla} \cdot (f^2{\bf \nabla} \phi),
\label{1}
\end{equation}
which is the equation of continuity,
\begin{equation}
\frac{\partial n}{\partial t}+{\bf \nabla} \cdot (n{\bf v})=0,
\label{111}
\end{equation}
and
\begin{equation}
-\hbar\frac{\partial \phi}{\partial t}=
-\frac{\hbar^2}{2mf} \nabla^2 f+\frac{1}{2}mv^2+V({\bf r})+U_0f^2.
\label{2}
\end{equation}
We  eliminate the phase variable by taking the gradient of (\ref{2}),
using ${\bf v}=\hbar{\bf \nabla}\phi/m$.
The resulting equation is written as
\begin{equation}
m\frac{\partial{\bf v}}{\partial t}=-{\bf\nabla}(\delta\mu+\frac{1}{2}mv^2),
\label{acc}
\end{equation}
where
\begin{equation}
\delta\mu=V+U_0n-\frac{\hbar^2}{2m\sqrt{n}} \nabla^2 \sqrt{n}-\mu_0.
\label{delta3}
\end{equation}
Since it is the gradient of $\delta\mu$, which enters the acceleration
equation (\ref{acc}), we are free to subtract a constant from
$\delta\mu$. We have  in (\ref{delta3}) chosen to subtract
the value of the equilibrium chemical potential $\mu_0$,
 which implies that  $\delta\mu$ is zero in equilibrium,
i.\ e.\ under stationary conditions.  The
equation $\delta\mu=0$ is  the time-independent
Gross-Pitaevskii equation.

In the Thomas-Fermi approximation,
the kinetic energy term is neglected, and the equilibrium density is therefore
given by
\begin{equation}
n_0U_0+V(x,y,z) =\mu_0.
\label{equi}
\end{equation}
Within the Thomas-Fermi approximation
one also neglects the kinetic energy term involving
$\delta n$ in the expression (\ref{delta3}) for $\delta\mu$. This yields
\begin{equation}
\delta\mu=U_0\delta n.
\label{deltamy}
\end{equation}
The Thomas-Fermi approximation for the modes should be a good one provided the
wavelength of the mode is large compared with the healing distance $\delta$.
With these approximations we may readily linearize the equations
(\ref{111})
 and (\ref{acc}), and eliminate $\delta\mu$ by means of
(\ref{deltamy}). The result is
\begin{equation}
m\frac{\partial^2\delta n}{\partial t^2}= U_0{\bf \nabla}
\cdot(n_0{\bf \nabla}\delta n).
\label{diff}
\end{equation}
If we only consider oscillations with the time dependence
 $\delta n\propto\exp(-i\omega t)$,
the differential equation (\ref{diff}) simplifies to
\begin{equation}
-\omega^2\delta n=\frac{U_0}{m}(
{\bf \nabla}n_0\cdot
{\bf \nabla}\delta n +
n_0\nabla^2\delta n).
\label{a}
\end{equation}

We investigate the surface modes in a two-dimensional configuration with the
linear ramp potential considered in the previous section.
In the $y$- and $z$-directions there is translational
invariance, and therefore the
solution must have the form of plane waves for these coordinates. We denote
the wavenumber of the mode by $q$, and take the direction of propagation to be
the
$z$-axis.
In the Thomas-Fermi approximation the condensate density  in equilibrium,
$n_0$,
is then given by $n_0(x)=-Fx/U_0$ for $x<0$, while it vanishes
for $x>0$. It follows that  (\ref{a}) has a solution of the form
\begin{equation}
\delta n = A e^{qx-iqz},
\label{tria1}
\end{equation}
which describes a wave propagating on the surface, and decaying
exponentially in the interior. Since (\ref{tria1})
satisfies $\nabla^2\delta n=0$, while the gradient
of the equilibrium density is given by $(-F/U_0,0,0)$,
we obtain by inserting (\ref{tria1}) into (\ref{a})
the dispersion relation
\begin{equation}
\omega^2 =\frac{F}{m}q.
\label{res1}
\end{equation}
This has  the same form as for a gravity wave propagating
on the surface of an incompressible ideal fluid in the
presence of a gravitational field $g=F/m$.

The solution (\ref{tria1}) is however not the only one which decays
exponentially in the interior. To investigate the solutions to
(\ref{a}) more generally we insert a function of the form
\begin{equation}
\delta n = f(qx) e^{qx+iqz},
\label{tria1a}
\end{equation}
and obtain the following second order differential equation for
$f(y)$,
\begin{equation}
y\frac{d^2f}{dy^2}+(2y+1)\frac{df}{dy}+(1-\epsilon)f=0,
\label{lag}
\end{equation}
where $\epsilon=\omega^2/gq$. By introducing the new variable $z=-2y$
one sees that Eq.\ (\ref{lag}) becomes the differential equation
for the Laguerre polynomials $L_n(z)$, provided $\epsilon - 1 = 2n$.
We have thus obtained the general dispersion
relation for the surface modes
\begin{equation}
\omega^2 =\frac{F}{m}q(1+2n), \;\;\;\; n=0,1,2\cdots
\label{res1a}
\end{equation}
with the associated density oscillations given by
\begin{equation}
\delta n(x,z,t)=AL_n(-2qx)e^{qx+iqz-i\omega t},
\end{equation}
with $A$ being an arbitrary constant.

To make contact with Stringari's calculation\cite{stringari},
we note that he found the
dispersion relation of modes in an isotropic harmonic trap to be given by
$\omega^2=\omega_0^2[l(1+2n)+3n +2n^2]$, where $l$ is the angular momentum
quantum number and $n$ the radial one, which gives the number of nodes in the
radial direction\cite{stringari}.  For $l$ much greater than 1, the
dispersion relation becomes  $\omega^2=\omega_0^2 l(1+2n)$.
  The wavenumber of the mode at the
surface of the cloud is given by $q=l/R$, and
therefore the dispersion relation
is $\omega^2=\omega_0^2 q R(1+2n)$, which is precisely the same as the result
(\ref{res1a}) we obtained above,
since the force due to the trap  at the
surface of the cloud is simply $F=\omega_0^2 R$ per
unit mass.  For large values of $l$ it is thus  a good
approximation to replace the
harmonic oscillator potential by the linear ramp,
as one might expect since the surface modes are
concentrated within a distance of order $R/l$ from the surface.
It should be noted that the  $n=0$  mode frequencies for the
plane surface with a linear ramp potential agree with the
frequencies  of the nodeless radial modes (corresponding to $n=0$)
for a harmonic trap at {\it all} values of $l$.
For modes with radial nodes ($n\neq 0$), the two results agree only when $l$
is much greater than $n$.

\section{Surface modes at shorter wavelengths }

When the healing length $\delta$ is not negligible compared with the
wavelength, there are corrections to the dispersion relation.
In the case of gravity waves on the surface of a liquid, modes at
shorter wavelengths are affected by the surface tension, and the dispersion
relation is given by $\omega^2=gq+\sigma q^3/\rho$, where
$\sigma$ is the surface tension, and $\rho$ is the
mass density of the fluid.  We now explore modes on the surface of a
Bose-Einstein condensed cloud at shorter wavelengths, and we shall show that
there are contributions to $\omega^2$ of order $q^4 \ln(1/q\delta)$
which may be
understood in terms of the effective surface tension introduced in the
discussion of static properties.

The basic problem is
to solve Eq.\ (\ref{acc}) including the quantum pressure
term in the expression for the chemical potential.  Rather than attacking the
problem directly, which leads to two coupled second-order differential
equations, we shall adopt a variational approach, which will allow us to
calculate the leading corrections to the Thomas-Fermi result for the mode
frequencies for small $q$.

In order to determine the dispersion relation
of surface modes at shorter wavelengths we
employ a trial wave function that allows us to calculate the total energy
in terms of two variables which describe the displacement of the surface
and the local velocity, respectively. In terms of these
variables the energy
functional assumes  the form of that of
a harmonic oscillator, from which we
may  extract the frequency as a function of $q$.

The trial wave function is motivated by the solution
found above in the Thomas-Fermi approximation. To lowest order
we may describe the motion of the surface in a traveling
wave by modifying the ground state wave function in two respects.  First
one shifts the spatial variable, thereby allowing for displacements of the
surface, and one introduces a phase factor to take into account motion of
the particles.  Explicitly, the wave function is given by
\begin{equation}
\psi(x,z,t)= \psi_{TF}(x-\Delta(x,z,t) )\exp{i\phi}
\approx \psi_{TF}(x) - \Delta(x,z,t)\psi'_{TF}(x) +i\phi\psi_{TF}(x).
\end{equation}
Here
$\Delta (x,z,t)=\xi_0 \exp qx \cos(qz -\omega t)$, with $\xi_0$
 constant, and
$ \phi=
(m v_0/\hbar q) e^{qx}  \sin(qz- \omega t)$,
 $v_{0}$ being the amplitude of the velocity at the surface of the cloud,
given by
\begin{equation}
{\bf v}=\frac{\hbar\nabla\phi}{m}=
v_0(\sin(qz-\omega t),0,\cos(qz-\omega t)).
\end{equation}

Let us now turn to the more general case.  We expect that
at frequencies small compared with the characteristic frequency $\sim
\hbar/2m\delta^2$ associated with adjustments of the density profile in the
region within $\sim \delta$ of the surface, the density profile will be
able to
adjust essentially
instantaneously to its equilibrium form corresponding to the
local number of
particles per unit area, even if the Thomas-Fermi approximation is not
valid. The real part of the wave function in the vicinity of the surface is
thus of the equilibrium form, but with a possible
translation perpendicular to the surface.
We shall therefore use a trial function which has the same form as for the
Thomas-Fermi case, but with the equilibrium Thomas-Fermi wave function
replaced by the exact one.  From this we shall calculate the energy, and
evaluate oscillation frequencies.  We write the wave function in the form
\begin{equation}
\psi = \psi_0 +\delta\psi,
\label{33}
\end{equation}
where $\delta\psi$ is the part due to the oscillation.
For the present purposes it is simplest to
consider a standing wave, and therefore we adopt the following form
\begin{equation}
\delta \psi(x,t) = (- \xi(t)  {\psi_{0}}'(x) +i \psi_{0}(x) \phi_0(t))
e^{qx} cos(qz).
\label{ans2}
\end{equation}
The $x$-component of the velocity of the surface is given either in terms
of the time derivative of the surface displacement, or in terms of the
$x$-derivative of the phase of the wave function.  This leads to the
consistency condition $\dot{\xi} =\hbar q\phi_0/m$.

We use the trial function (\ref{ans2}) to evaluate the energy functional
\begin{equation}
E= \int d {\bf r} \left[
{\hbar^2 \over 2m} {|\nabla \psi|^2 } +
V({\bf r}) |\psi({\bf r},t)|^2 +
\frac{1}{2}U_{0} |\psi({\bf r},t)|^4 \right],
\label{funct}
\end{equation}
to second order in $\xi$ and $\dot\xi$.
The zeroth order term gives rise to an unimportant constant, while
the first order term vanishes, since the
trial function is a solution of the Gross-Pitaevskii equation.
The interesting physics is contained in the second order term
$E^{(2)}$. The $\xi^2$-part of the kinetic-energy contribution contains
an integral over $x$ of the form
\begin{equation}
\int_{-\infty}^{\infty} dx {({\psi_0}'e^{qx})}'{({\psi_0}'e^{qx})}'=
-\int_{-\infty}^{\infty}
dx e^{2qx}\psi_0'({\psi_0}''' + 2q{\psi_0}''+ q^2{\psi_0}').
\label{differ}
\end{equation}
The third-order derivative in (\ref{differ}) is eliminated by
differentiating the Gross-Pitaevskii equation,
\begin{equation}
-\frac{\hbar^2}{2m}\psi_0'''=-{V}'\psi_0-V{\psi_0}'-3U_0\psi_0^2{\psi_0}'.
\label{gp3}
\end{equation}
The last two terms on the right hand side of (\ref{gp3}) yield
contributions that are canceled by those coming from
the potential and interaction energies in (\ref{funct}).
The remaining terms may be combined using $V'=F$ and partial
integration, and yield for the energy per unit area
\begin{equation}
E^{(2)}  ={\xi^2\over 4}
{\left[Fq \int^{\infty}_{-\infty} dx
  e^{2qx} {\psi_0}^2+
 {\hbar^2 \over m}
\int^{\infty}_{-\infty} dx \left(q^2 e^{2qx} \psi_{0}'^2\right)
  \right]}
+  {m {\dot\xi}^2 \over 4 }
\int^{\infty}_{-\infty} dx e^{2qx}{\psi_{0}}^2 .
\end{equation}
This is of the same form  as for an harmonic oscillator,
$ E^{(2)}={1 \over 2} C_{1} \xi^2 + {1 \over 2} C_{2} {\dot\xi}^2 $
where $C_{1}$ and $ C_{2} $ are constants, and
the frequency is given by $\omega^2=C_1/C_2$. Integrating by parts
the term involving $F$, we obtain the final result
\begin{equation}
\omega^2 =  {F \over m} q +
{\hbar^2 q^4 \over m^2} I(q),
\label{grav2}
\end{equation}
where the dimensionless quantity $I(q)$ is given by
\begin{equation}
I(q)=
\frac{\int^{\infty}_{-\infty} dx
 e^{2qx} ({\psi_{0}}')^2  }{q^2\int^{\infty}_{-\infty} dx
  e^{2qx} {\psi_{0}}^2  }.
\label{grav3}
\end{equation}
The first term in (\ref{grav2}) gives the frequency of the surface mode in
the Thomas-Fermi approximation, while the second term involving $I(q)$ is
a correction term.

We may evaluate the leading long-wavelength corrections to the
dispersion relation $I(q)$ for
$q\ll
1/\delta$ by splitting up the range of integration into two regions
\begin{equation}
\int_{-\infty}^{\infty}dxe^{2qx}({\psi_{0}}')^2=
\int_{-\infty}^{-L}dxe^{2qx}({\psi_{0}}')^2+
\int_{-L}^{\infty}dxe^{2qx}({\psi_{0}}')^2,
\end{equation}
 where $\delta\ll L\ll 1/q$. In the first of the two
integrals  we may use the Thomas-Fermi approximation
${\psi_0} =\sqrt{-Fx/U_0}$, since $L\gg \delta$.
In the second we may replace $\exp 2qx$
by 1, since $ql\ll 1$.
By using partial integration one finds
\begin{equation}
\int_{-L}^{\infty}dx({\psi_{0}}')^2=\frac{F}{2U_0}
- \int_{-L}^{\infty}dx{\psi_{0}}'' {\psi_{0}}
\label{kinen}
\end{equation}
since the value of $\psi_0\psi_{0}'$ at $x=-L$
may be evaluated in the Thomas-Fermi approximation, resulting in $-F/2U_0$.
The integral
on the right hand side of  (\ref{kinen}) occurs in calculations of the
kinetic energy associated with the surface region, and it has been
evaluated previously.  Its asymptotic form is $(F/4U_0)\ln(L/1.776\delta)$,
which corresponds to the kinetic energy contribution in Thomas-Fermi theory
cut off at a distance $1.776\delta$, where the coefficient was determined
by numerical integration.
 We obtain consequently
\begin{equation}
\int_{-\infty}^{-l}dxe^{2qx}({\psi_{0}}')^2=
\frac{F}{4U_0}(2-\ln 2qL-\ln \gamma)+\frac{F}{4U_0}\ln\frac{L}{1.776
\delta},
\end{equation}
where $\gamma \;(\approx 1.778)$ is the Euler constant.
The denominator in (\ref{grav3}) is evaluated in the
Thomas-Fermi approximation, resulting in $F/4U_0$.
We finally obtain
\begin{equation}
I(q)\simeq -\ln q\delta  +\ln(e^2/3.552\gamma)\simeq -\ln q\delta +0.15.
\end{equation}
Thus we find
\begin{equation}
\omega^2 \approx  {F \over m} q +
{\hbar^2 q^4 \over m^2} \left[ -\ln q\delta +0.15\right] .
\label{final}
\end{equation}
The qualitative behavior is easy to understand by analogy with the surface
tension contribution to the frequency of gravity waves on the surface of a
fluid, where the contribution to $\omega^2$ is $\sigma q^3/\rho$.  In the
present problem, the effective surface tension depends logarithmically on the
length scale, which is given by the wavelength.
The $q^4$-dependence exhibited by (\ref{final})
then results  from dividing $q^3$ by the  effective density in the
region where the fluid is moving, this being of order
the fluid density at the distance  $1/q$ from the surface, or $F/U_0q$.
Thus one sees that the surface
tension is very weakly dependent on the trap parameters and the atomic
scattering length, which occur only in the logarithm.

This result agrees to the order indicated with the result of
using the full solution $\psi_0(x)$
to the (equilibrium) Gross-Pitaevskii equation in evaluating the
expression (\ref{grav3}).               In a recent paper,
Fetter and Feder\cite{FF}
analyzed the corrections to the excitation frequencies in  the
Thomas-Fermi limit\cite{stringari} for an
atomic cloud  confined by a
spherically symmetric trap. By employing the matching
conditions of boundary-layer theory\cite{BO}, they were able to demonstrate
 that the leading correction to the Thomas-Fermi limit is of
order $R^{-4}$, where $R$ is the  Thomas-Fermi  radius,
while terms of order $R^{-4}\ln R$ were found to be absent.
Repeating their analysis for the different geometry which we are
considering, we have explicitly verified that
the  matching conditions  allow for the presence of
 terms of order $q^4\ln q$, which we have
found in the present paper.

This result for the surface mode frequency can be obtained
in a more rigorous fashion by a variational approach.
The resulting equations of motion of $\xi$ and $\phi_{0}$ take
the form of that of a classical harmonic oscillator with frequency
given by (\ref{grav2}). The details of this calculation are described in
Appendix A.

\vspace{0.5cm}

\noindent We thank Emil Lundh for helpful contributions.

\appendix
\section{variational approach}
The Gross-Pitaevskii equation may be derived from the variational principle
\begin{equation}
\delta \int dt L=0,
\end{equation} 
where
\begin{equation}
L=\int d{\bf r}
{i\hbar\over2}\left(
\psi^{*}{\partial\psi\over\partial t}-
\psi{\partial\psi^{*}\over\partial t}
\right)-E.
\end{equation}
Here $E$ is the energy functional given by equation (\ref{funct}).
For $\psi$ we adopt our ansatz, (\ref{33}), and the equations of
motion for $\xi$
and $\phi_{0}$ may then be determined from the variational principle.
To second order in $\xi$ and $\phi_{0}$ the Lagrangian is
\begin{equation}
L={\hbar\over4}I_{1}(\xi{\dot \phi_{0}}-\phi_{0}{\dot \xi})- 
(E^{(0)}+E^{(2)}),
\end{equation}
where $E^{(0)}$ is the ground-state energy obtained by inserting
$\psi=\psi_{0}$
in equation (\ref{funct}), while $E^{(2)}$ is given by
\begin{equation}
E^{(2)}={\xi^{2}\over4}
\left[
Fq I_{2}+{\hbar^{2}\over m}I_{3}
\right]
+{\hbar^{2} q^{2}\over4m}\phi_{0}^{2}I_{2},
\end{equation}
in terms of the integrals
\begin{equation}
I_{1}={1\over2}\int_{-\infty}^{\infty}dx e^{2qx}{(\psi_{0}^{2})}^{'},
\label{i1}
\end{equation}
\begin{equation}
I_{2}=\int_{-\infty}^{\infty}dx e^{2qx}\psi_{0}^{2},
\end{equation}
and
\begin{equation}
I_{3}=\int_{-\infty}^{\infty}dx (q^{2}e^{2qx}{\psi_{0}^{'}}^{2}).
\end{equation}
The resulting equations of motion for $\xi$ and $\phi_{0}$ are
\begin{equation}
{\hbar\over 2}I_{1}{\dot\phi_{0}}-
{1\over2}\xi(FqI_{2}+
{\hbar^{2}\over m}I_{3})=0
\end{equation}
and
\begin{equation}
{\hbar\over2}I_{1}{\dot \xi}+{\hbar^{2}q^{2}\over2m}
I_{2}\phi_{0}=0.
\end{equation}
By partial integration of (\ref{i1}) we see that $I_{1}=-qI_{2}$, and
subsequently, by eliminating $\phi_{0}$ between the last two equations,
one obtains the result (\ref{grav2}).


\begin{thebibliography} {99}


    \bibitem{123} M.~H.~Anderson, J.~R.~Ensher, M.~R.~Matthews,
C.~E.~Wieman,
and E.~A.~Cornell, Science {\bf 269}, 198 (1995).
 C.~C.~Bradley, C.~A.~Sackett, J.~J.~Tollett, and R.~G.~Hulet,
Phys.\ Rev.\ Lett.\ {\bf 75}, 1687 (1995).
K.~B.\ Davis, M.-O.\ Mewes, M.~R.\ Andrews, N.~J.\ van Druten,
D.~S.\ Durfee, D.~M.\ Kurn, and W.\ Ketterle, Phys.\ Rev.\ Lett.\ {\bf 75},
3969 (1995).

\bibitem{ruckenstein leggett bp}
V.\ V.\ Goldman, I.\ F.\ Silvera, and
A.\ J.\ Leggett, Phys.\ Rev.\ B{\bf 24}, 2870 (1981);
G.\ Baym and C.\ J.\ Pethick,
Phys.\ Rev.\ Lett.\  {\bf 76}, 6 (1996).


    \bibitem{stringari} S.\ Stringari,
Phys.\ Rev.\ Lett.\  {\bf 77}, 2360 (1996).


    
\bibitem{pitaevskii lundh fetter}F.\ Dalfovo, L.\ P.\
Pitaevskii, and S.\ Stringari, Phys.\ Rev.\ A {\bf 54},
4213 (1996);
 E.~Lundh, C.~J.~Pethick, and H.~Smith, Phys.~Rev.~A {\bf 55},
 2126 (1997).

\bibitem{lps} E.~Lundh, C.~J.~Pethick, and H.~Smith, Phys.~Rev.~A {\bf 55},
2126 (1997).


\bibitem{FF} A.\ L.\ Fetter and D.\ L.\ Feder, cond-mat/9704173.

\bibitem{BO} See, e.\ g., C.\ M.\ Bender and S.\ A.\ Orszag,
Advanced Mathematical Methods for
Scientists and Engineers (McGraw Hill, NY, 1978), Chap.\ 9.


\end{thebibliography}
\end{document}